\documentclass[%
 aip,
 amsmath,amssymb,
 reprint,%
]{revtex4-1}

\usepackage{graphicx}
\usepackage{dcolumn}
\usepackage{bm}
\usepackage{physics}

\usepackage[utf8]{inputenc}
\usepackage[T1]{fontenc}
\usepackage{mathptmx}

\usepackage{etoolbox}
\usepackage[nameinlink]{cleveref}

\crefname{figure}{Fig.}{Figs.}
\Crefname{figure}{Figure}{Figures}

\crefname{equation}{Eq.}{Eqs.}
\Crefname{equation}{Equation}{Equations}

\makeatletter
\def\@email#1#2{%
 \endgroup
 \patchcmd{\titleblock@produce}
  {\frontmatter@RRAPformat}
  {\frontmatter@RRAPformat{\produce@RRAP{*#1\href{mailto:#2}{#2}}}\frontmatter@RRAPformat}
  {}{}
}%
\makeatother
\begin{document}

\preprint{AIP/123-QED}

\title{Spin-coherence characterization of boron vacancy defects in hexagonal boron nitride with broadband microwave pulses}

\author{Yuki Nakamura}
\affiliation{ 
Department of Physics, The University of Tokyo, 7-3-1 Hongo, Bunkyo, Tokyo 113-0033, Japan
}%

\author{Takuya Iwasaki}
\affiliation{%
Research Center for Materials Nanoarchitectonics, National Institute for Materials Science, 1-1 Namiki, Tsukuba, Ibaraki 305-0044, Japan
}%

\author{Shu Nakaharai}
\affiliation{%
Department of Electric and Electronic Engineering, Tokyo University of Technology, 1404-1 Katakuramachi, Hachiohji, Tokyo 192-0982, Japan
}%

\author{Shinichi Ogawa}
\author{Yukinori Morita}
\affiliation{%
 National Institute of Advanced Industrial Science and Technology, 1-1-1 Umezono, Tsukuba, Ibaraki 305-8568, Japan
}%

\author{Kenji Watanabe}
\affiliation{%
Research Center for Electronic and Optical Materials, National Institute for Materials Science, 1-1 Namiki, Tsukuba, Ibaraki 305-0044, Japan
}%

\author{Takashi Taniguchi}
\affiliation{%
Research Center for Materials Nanoarchitectonics, National Institute for Materials Science, 1-1 Namiki, Tsukuba, Ibaraki 305-0044, Japan
}%

\author{Kento Sasaki}
\affiliation{ 
Department of Physics, The University of Tokyo, 7-3-1 Hongo, Bunkyo, Tokyo 113-0033, Japan
}%

\author{Kensuke Kobayashi}
\affiliation{ 
Department of Physics, The University of Tokyo, 7-3-1 Hongo, Bunkyo, Tokyo 113-0033, Japan
}%
\affiliation{%
Trans-scale Quantum Science Institute, The University of Tokyo, 7-3-1 Hongo, Bunkyo, Tokyo 113-0033, Japan}%

\date{\today}

\begin{abstract}
Negatively charged boron vacancy ($\mathrm{V_{B}^-}$) defects in hexagonal boron nitride (hBN) are promising for nanoscale-proximity quantum sensing.
To evaluate their performance, it is important to characterize the spin coherence times $T_2^*$ and $T_2$.
In this study, we realized sub-GHz Rabi oscillations of $\mathrm{V_{B}^-}$ using an isotopically enriched $\mathrm{h}^{10}\mathrm{B}^{15}\mathrm{N}$ thin film directly stamped onto a narrow gold wire.
Using these strong microwave pulses, we performed Ramsey interference and Hahn echo measurements.
The Ramsey interference signal showed Gaussian-like decay, yielding $T_2^* = 13.8 \pm 0.5$ ns.
The Hahn echo measurement gave $T_2 = 108.7 \pm 5.5$ ns and a stretch factor of $\alpha = 1.25 \pm 0.11$.
These results experimentally clarify the spin coherence properties of $\mathrm{V_{B}^-}$ and provide an effective method for evaluating the coherence of spin defects in van der Waals thin films with broad resonance linewidths.

\end{abstract}

\maketitle
Optically addressable spin defects in wide-bandgap semiconductors have been widely studied as platforms for quantum sensing and quantum information processing~\cite{weber2010quantum,atature2018material,awschalom2018quantum,wolfowicz2021quantum}.
Negatively charged boron vacancy ($\mathrm{V_{B}^-}$) defects in hexagonal boron nitride (hBN) are optically addressable spin defects found in a van der Waals (vdW) material~\cite{gottscholl2020initialization}.
Because an hBN thin film can be placed in direct contact with the target surface, these defects are promising for quantum sensing that brings the spin defect within a few nanometers of the sample~\cite{durand2023optically,zhou2024sensing,kumar2022magnetic,healey2023quantum,huang2022wide,zhou2024sensing}.
For these applications, the coherence of the $\mathrm{V_{B}^-}$ electron spin is an important quantity that determines device performance.
In particular, spin-coherence time $T_2^*$ and $T_2$ determine how long the quantum state can be used, and the stretch factor of the decay provides information on the decoherence mechanism~\cite{Merkulov2002,Dobrovitski2008,Maze2008PRB,Bauch2020}.

To measure the coherence properties of the $\mathrm{V_{B}^-}$ electron spin accurately, it is important to excite the entire resonance spectrum.
In $\mathrm{V_{B}^-}$, the hyperfine interaction with nearby nuclear spins is strong~\cite{haykal2022decoherence,Trknyi2025}, so that the resonance lines split by the nearest nitrogen nuclear spins are additionally broadened by the inhomogeneous hyperfine fields from nearby nuclear spins and overlap with one another, giving an effective resonance linewidth of several hundred MHz~\cite{haykal2022decoherence,gu2023multi,clua2023isotopic,gong2024isotope}.
If one uses a microwave power that is sufficient to excite a single hyperfine component, for example a Rabi frequency of $f_\mathrm{R} \sim 80$~MHz, pulse errors are suppressed for that component, but neighboring hyperfine components are only partially excited because of detuning.
As a result, the measured signal contains both on-resonance components with small pulse errors and off-resonance components with significant pulse errors, which can systematically bias the extracted coherence times.
To avoid this problem and to evaluate $T_2^*$ and $T_2$ accurately, one needs microwave pulses with a Rabi frequency in the sub-GHz range, high enough to cover the full spectral range of the hyperfine splitting.

Coherence measurements of the $\mathrm{V_{B}^-}$ electron spin have been reported previously.
For Hahn echo, some studies estimated $T_2\sim 2$--$15~\mathrm{\mu s}$ with $f_\mathrm{R}=10$--20~MHz~\cite{gottscholl2021room,Liu2022,Murzakhanov2022}, while others reported $T_2\sim 70~\mathrm{ns}$ with $f_\mathrm{R}=20$--83~MHz~\cite{haykal2022decoherence,Rizzato2023,Gong2023,gong2024isotope}.
A value of $T_2\sim 186~\mathrm{ns}$ has also been reported in isotopically enriched $\mathrm{h}^{10}\mathrm{B}^{15}\mathrm{N}$~\cite{gong2024isotope}.
For Ramsey interference, a previous study reported $T_2^* \sim 1~\mathrm{\mu s}$ under the condition $f_\mathrm{R}<20$~MHz~\cite{Liu2022}.
The wide spread of reported coherence times suggests that pulse errors play an important role in their estimation.

In this Letter, we evaluate $T_2^*$ and $T_2$ under conditions where pulse errors are suppressed by applying strong microwave pulses with $f_\mathrm{R} =$~0.35--0.6~GHz to $\mathrm{V_{B}^-}$ in isotopically enriched $\mathrm{h}^{10}\mathrm{B}^{15}\mathrm{N}$~\cite{clua2023isotopic,gong2024isotope}.
Using a device in which an hBN thin film was directly stamped onto a narrow gold wire~\cite{Nakamura2025}, we realized fast Rabi oscillations with $f_\mathrm{R} = 421$~MHz.
From Ramsey interference measured with microwave pulses at $f_\mathrm{R} = 600$~MHz, we obtain $T_2^* = 13.8 \pm 0.5$~ns with a stretch factor of $\alpha = 2$.
From Hahn echo measured with microwave pulses at $f_\mathrm{R} = 357$~MHz, we obtain $T_2=108.7\pm5.5$~ns and a stretch factor of $\alpha=1.25\pm0.11$.
These results are important because they provide a more reliable evaluation of the coherence of the $\mathrm{V_{B}^-}$ electron spin by using broadband excitation.

\begin{figure}[t]
    \centering
    \includegraphics[width=\columnwidth,pagebox=cropbox]{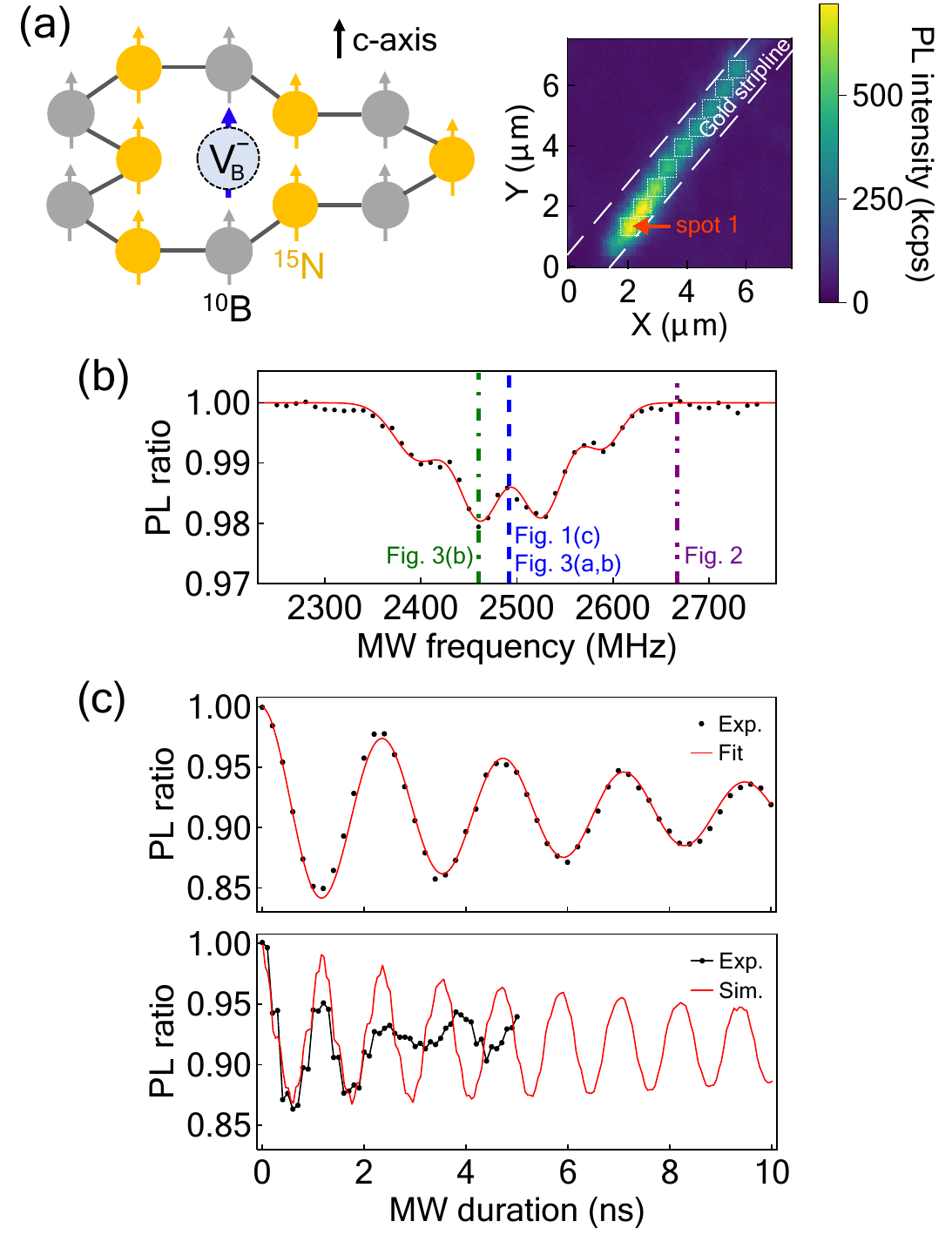}
    \caption{
    (a)~[Left panel]~Defect structure of $\mathrm{V_{B}^-}$ in $\mathrm{h}^{10}\mathrm{B}^{15}\mathrm{N}$.
    [Right panel]~Confocal microscope image of the $\mathrm{V_{B}^-}$ spots created in the hBN thin film.
    All measurements were performed on ``spot 1''.
    (b)~ODMR spectrum of the $m_S = 0 \leftrightarrow -1$ transition measured with $B_z=34.5$~mT applied parallel to the crystal c-axis of hBN.
    The vertical axis shows the PL intensity normalized by that measured with the microwave off, and the horizontal axis is the microwave frequency $f_\mathrm{mw}$.
    The blue dashed line corresponds to $f_\mathrm{mw}=2492$~MHz, the green dashed line to $f_\mathrm{mw}=2460$~MHz, and the purple dashed line to $f_\mathrm{mw}=2667$~MHz.
    (c)~Experimental results of Rabi oscillations.
    The microwave frequency was set to $f_\mathrm{mw}=2492$~MHz [the blue dashed line in \cref{fig:VBpulse_Basic}(b)].
    The horizontal axis is the microwave pulse duration.
    [Upper panel]~Rabi oscillation with a Rabi frequency of $f_\mathrm{R}=421\pm1$~MHz.
    [Lower panel]~Coherent oscillation observed when the microwave power was set to three times that in the upper panel.
    }
    \label{fig:VBpulse_Basic}
\end{figure}

\begin{figure}
    \centering
    \includegraphics[width=\linewidth]{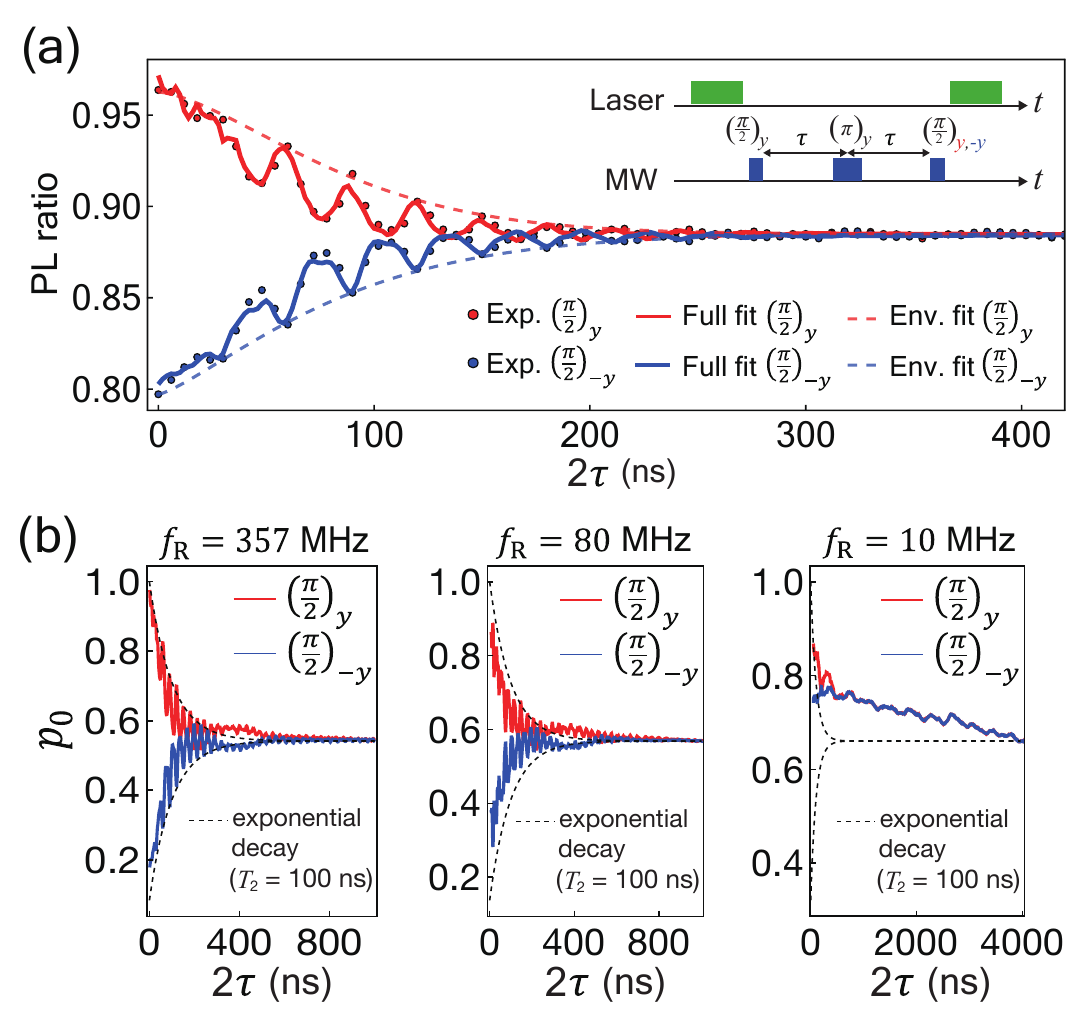}
    \caption{
    (a)~Experimental result of Hahn echo measured at $f_\mathrm{R} = 357$ MHz and $f_\mathrm{mw} = 2492$ MHz.
    The dashed line shows the envelope fit based on \cref{eq:fitformula_VB_Hahn_env}, and the solid line shows the full-waveform fit based on \cref{eq:fitformula_VB_Hahn}.
    (b)~Simulation of Hahn echo.
    The Lindblad equation was numerically solved with $T_2=100$~ns and $\theta = 3.8^\circ$, while all other parameters were set to the same values as in a previous study~\cite{Nakamura2025}.
    The vertical axis is the probability $p_0$ of observing $m_S=0$, and the horizontal axis is the total waiting time $2\tau$.
    [Left panel]~$f_\mathrm{R} = 357$ MHz and $f_\mathrm{mw} = 2492$ MHz, [middle panel]~$f_\mathrm{R} = 80$ MHz and $f_\mathrm{mw} = 2460$ MHz, and [right panel]~$f_\mathrm{R} = 10$ MHz and $f_\mathrm{mw} = 2460$ MHz.
    }
    \label{fig:VBpulse_HahnEcho}
\end{figure}

We use a narrow gold wire with width $w=1.5~\mathrm{\mu m}$, which enables the application of a strong microwave magnetic field to $\mathrm{V_{B}^-}$~\cite{Nakamura2025}.
Compared with the gold wires with width $w=50~\mathrm{\mu m}$ that have been widely used in previous studies~\cite{Liu2022,Gong2023,gong2024isotope}, our $w=1.5~\mathrm{\mu m}$ wire is expected to provide more than ten times stronger microwave driving~[See Supplementary].
We measured the electron spin of $\mathrm{V_{B}^-}$ in isotopically enriched $\mathrm{h}^{10}\mathrm{B}^{15}\mathrm{N}$ synthesized under high-pressure and high-temperature conditions~\cite{sasaki2023nitrogen}.
An $\mathrm{h}^{10}\mathrm{B}^{15}\mathrm{N}$ flake with thickness below 30~nm was stamped onto a gold wire with thickness 100~nm and width $w=1.5~\mathrm{\mu m}$, fabricated by electron-beam lithography (Elionix ELS-F125).
The flake on the wire was then irradiated with $\mathrm{He}^{+}$ ions using a helium ion microscope at a dose of $10^{16}~\mathrm{/cm^2}$ and an acceleration voltage of 30~keV, thereby creating a $(300)^2~\mathrm{nm^2}$ spot consisting of a $\mathrm{V_{B}^-}$ ensemble~[\cref{fig:VBpulse_Basic}(a), right panel]~\cite{sasaki2023magnetic,gu2024systematic}.
In all experiments, a magnetic field of $B_z=34.5$~mT was applied to this $\mathrm{V_{B}^-}$ spot along the crystal c-axis.
As will be discussed later in detail, the misalignment with respect to the c-axis was estimated to be $\theta=3.8^\circ$ from the Hahn echo analysis in \cref{fig:VBpulse_HahnEcho}(a).

\begin{figure}
    \centering
    \includegraphics[width=\columnwidth,pagebox=cropbox]{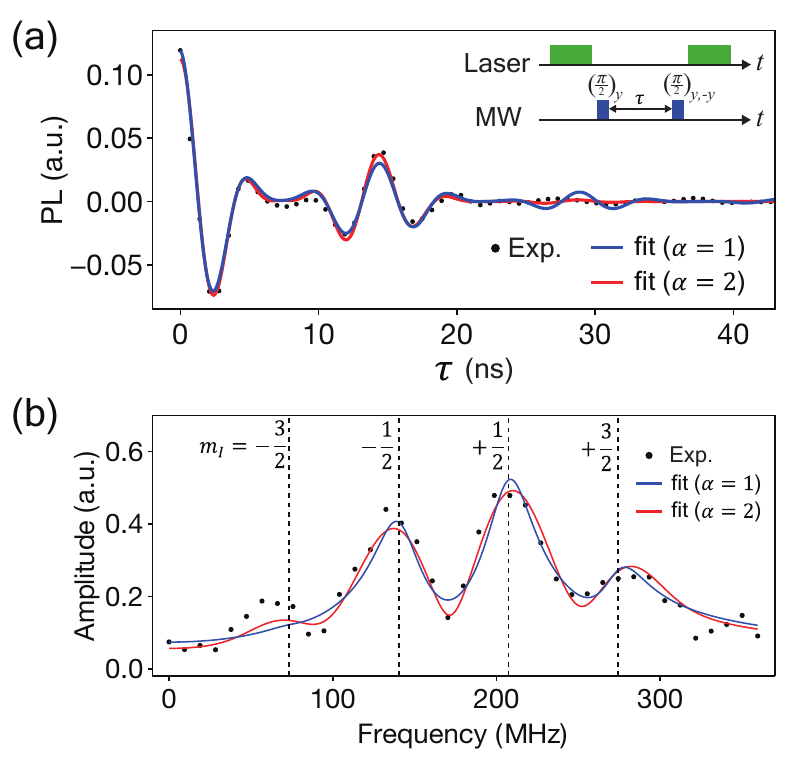}
    \caption{
    (a)~Experimental result of Ramsey interference.
    The microwave frequency was set to $f_\mathrm{mw}=2667$~MHz [the purple dashed line in \cref{fig:VBpulse_Basic}(b)].
    The vertical axis is the PL intensity, and the horizontal axis is the waiting time $\tau$.
    (b)~Fourier transform spectrum of (a).
    The four vertical black dashed lines indicate the frequency components $f_n$ ($n=\{1,2,3,4\}$) of the Ramsey fringes, which correspond to the hyperfine splitting due to three neighboring $^{15}\mathrm{N}$ nuclear spins.
    }
    \label{fig:VBpulse_Ramsey}
\end{figure}

The ODMR spectrum is shown in \cref{fig:VBpulse_Basic}(b).
Four dips arising from the hyperfine interaction with three neighboring $^{15}\mathrm{N}$ nuclear spins are observed, showing a typical spectral shape in which the four components overlap with each other.
In all experiments except for the Ramsey interference shown in \cref{fig:VBpulse_Ramsey}, the microwave frequency was set to the center of the resonance spectrum, $f_\mathrm{mw}=2492$~MHz, corresponding to a detuning of $\delta=0$~MHz [blue dashed line].
By contrast, in the Ramsey interference shown in \cref{fig:VBpulse_Ramsey}, $T_2^*$ of $\mathrm{V_B^-}$ is short, so a large detuning is required in order to observe a sufficient number of fringes within that timescale.
We therefore set the microwave frequency to $f_\mathrm{mw}=2667$~MHz, corresponding to $\delta=175$~MHz.
In the Hahn echo simulations shown in \cref{fig:VBpulse_HahnEcho}(b), the microwave frequency was set to $f_\mathrm{mw}=2492$~MHz, corresponding to $\delta=0$~MHz, for the case of strong microwave driving [left panel].
By contrast, for weak microwave driving [middle and right panels], the microwave frequency was set to $f_\mathrm{mw}=2460$~MHz, corresponding to $\delta=-32$~MHz, in order to excite one of the four hyperfine components selectively and efficiently.

We first show the acceleration of the Rabi oscillation enabled by the narrow wire.
As shown in the upper panel of \cref{fig:VBpulse_Basic}(c), we observed sub-GHz Rabi oscillations.
By fitting the data with
\begin{equation}
I(\tau)
= \left(1-\dfrac{C}{2}\right)+\dfrac{C}{2}\exp\left\{-\left(\dfrac{\tau}{T_\mathrm{2,Rabi}}\right)^\alpha\right\}\cos{(2\pi f_\mathrm{R} \tau)},
\label{eq:fitformula_VB_Rabi_decay} 
\end{equation}
we obtained $f_\mathrm{R}=421\pm1$~MHz, $C=17.3\pm0.2$~\%, $T_\mathrm{2,Rabi}=7.3\pm0.1$~ns, and $\alpha=0.90\pm0.12$, which gives a $\pi$ pulse with duration $t_\pi \equiv 1/(2 f_\mathrm{R})=1.2$~ns.

As shown in the lower panel of \cref{fig:VBpulse_Basic}(c), when the microwave amplitude is set to three times that in the upper panel, coherent dynamics on the GHz scale is observed.
However, in this case, $f_\mathrm{R}=3\times421~\mathrm{MHz}=1.26$~GHz, so the condition for the rotating-wave approximation, $f_\mathrm{R} \ll f_{\mathrm{mw}}$, is not satisfied.
Therefore, this oscillation is not a single sinusoidal oscillation, but is likely a more complex dynamics described by Floquet theory~\cite{Fuchs2009}.
If a stronger static magnetic field is applied, Rabi oscillations above 1~GHz may also be realized by setting the microwave frequency to $f_\mathrm{mw} \gtrsim 8$~GHz.

Next, we performed Ramsey interference measurements under strong microwave driving.
The microwave frequency was set to $f_\mathrm{mw}=2667$~MHz [the purple dashed line in \cref{fig:VBpulse_Basic}(b)], and the $\pi/2$ pulse width was set to $t_{\pi/2}=0.4$~ns.
\Cref{fig:VBpulse_Ramsey}(a) shows the dependence on the waiting time $\tau$ of the difference in photoluminescence (PL) intensity obtained by switching the relative phase of the final $\pi/2$ pulse between $0^\circ$ and $180^\circ$.
The Fourier transform spectrum is shown in \cref{fig:VBpulse_Ramsey}(b).

The four vertical black dashed lines in \cref{fig:VBpulse_Ramsey}(b) represent the frequency components of the Ramsey fringes arising from the hyperfine interaction with three neighboring $^{15}\mathrm{N}$ nuclear spins,
\begin{equation}
f_n=\delta+m_I^{(n)}A_{zz},
\label{eq:VB_RamseyFringe}
\end{equation}
where $m_I^{(n)}=\left\{-\frac{3}{2},-\frac{1}{2},+\frac{1}{2},+\frac{3}{2}\right\}$, $n=\{1,2,3,4\}$, and $A_{zz}=-67$~MHz.
The four peaks observed in \cref{fig:VBpulse_Ramsey}(b) are consistent with the four hyperfine-split components seen in the ODMR spectrum in \cref{fig:VBpulse_Basic}(b), and these components are clearly separated from each other.
Therefore, in this measurement, the inhomogeneous broadening intrinsic to the $\mathrm{V_B^-}$ electron spin can be evaluated after separating the known hyperfine splittings due to the three neighboring $^{15}\mathrm{N}$ nuclear spins.

To quantify this inhomogeneous broadening, we fitted the experimental data in \cref{fig:VBpulse_Ramsey}(a) with
\begin{equation}
I(\tau)
=
\exp\left\{-\left(\frac{\tau}{T_2^*}\right)^\alpha\right\}
\sum_{n=1}^{4} C_n \cos(2\pi f_n \tau)
\label{eq:fitformula_VB_Ramsey}
\end{equation}
with fitting parameters $\{C_1,C_2,C_3,C_4,T_2^*,\delta,\alpha\}$.
By comparing fits with $\alpha=2$ and $\alpha=1$ fixed, we obtained $T_2^*=13.8\pm0.5$~ns and $10.6\pm0.7$~ns, respectively.
The corresponding mean squared errors (MSEs) were $1.25\times10^{-5}$ for $\alpha=1$ and $9.65\times10^{-6}$ for $\alpha=2$, showing that the assumption $\alpha=2$ describes the experimental data better.
The case $\alpha=2$ corresponds to a Gaussian inhomogeneous distribution~\cite{slichter2013principles}, and the full width at half maximum estimated from $T_2^*=13.8$~ns is $\frac{2\sqrt{\ln2}}{\pi T_2^*}=38$~MHz.

This full width at half maximum is narrower than the 44--55~MHz estimated from the ODMR spectra of $\mathrm{V_B^-}$ in $\mathrm{h}^{10}\mathrm{B}^{15}\mathrm{N}$~\cite{clua2023isotopic,gong2024isotope}.
This suggests that avoiding power broadening is important for evaluating the inhomogeneous broadening.
In addition, recent theoretical studies have shown that the decoherence of the $\mathrm{V_B^-}$ electron spin strongly depends on the strong Fermi-contact interaction with several tens of nearby nuclear spins~\cite{haykal2022decoherence,Trknyi2025}.
The Gaussian-like inhomogeneous distribution obtained in the present measurement, where the effect of power broadening is suppressed, suggests that the effective local magnetic-field distribution produced by the hyperfine interaction with such nearby nuclear spins may also be Gaussian-like.

We further measured Hahn echo at a microwave frequency of $f_\mathrm{mw}=2492$~MHz and a Rabi frequency of $f_\mathrm{R}=357$~MHz.
As shown in \cref{fig:VBpulse_HahnEcho}(a), a Hahn echo signal with clear oscillations was observed, and its envelope decayed on a timescale of $T_2\sim100$~ns.
This decay time is broadly consistent with the values reported in previous studies using $f_\mathrm{R}=20$--83~MHz~\cite{haykal2022decoherence,Rizzato2023,Gong2023,gong2024isotope}.
The frequency of the observed oscillation is close to $|A_{zz}|=67$~MHz, indicating that it is electron spin echo envelope modulation (ESEEM) arising from the hyperfine interaction with three neighboring $^{15}\mathrm{N}$ nuclear spins~\cite{Rizzato2023}.
Although the clarity of the ESEEM and the decay timescale differ among previous studies, these features are considered to depend strongly on the strength of the microwave pulses.
In the following, we examine this point on the basis of simulation.

\Cref{fig:VBpulse_HahnEcho}(b) shows the simulation results based on the Lindblad equation for a $\mathrm{V_B^-}$ electron spin coupled to three neighboring $^{15}\mathrm{N}$ nuclear spins~[See Supplementary]~\cite{Nakamura2025}.
In this simulation, the decoherence term was set to a relaxation rate corresponding to $T_2=100$~ns.
Therefore, if the pulses are sufficiently ideal, the decay of the envelope of the transition probability $p_0$ is expected to appear on a similar timescale.
However, at $f_\mathrm{R}=10$~MHz, the apparent decay of the envelope is found to be distorted to a much longer timescale of several $\mathrm{\mu s}$ or more.
This behavior closely reproduces the Hahn echo signals reported in previous studies that used $f_\mathrm{R}=10$--20~MHz and estimated $T_2\sim 2$--$15~\mathrm{\mu s}$~\cite{gottscholl2021room,Liu2022,Murzakhanov2022}.
Thus, under weak driving, the pulses become long, and decoherence during the driving and off-resonant errors become significant, so the evaluation of $T_2$ loses quantitative reliability.
At $f_\mathrm{R}=80$~MHz, by contrast, the envelope decay time is roughly consistent with $T_2$, but the signal at short $2\tau$ becomes complicated because on-resonant and off-resonant components are mixed.
In contrast, at $f_\mathrm{R}=357$~MHz, the Rabi frequency exceeds the detuning caused by the hyperfine interaction, so nearly identical $\pi/2$ and $\pi$ pulses are realized for all hyperfine components, and the effects of pulse errors and time evolution during the driving are suppressed.
As a result, the dominant ESEEM component is observed more clearly, which stabilizes the analysis of the Hahn echo signal.

Next, we describe how the coherence time $T_2$ and the stretch factor $\alpha$ were estimated.
In the simulation, the polar angle $\theta$ between the magnetic field and the c-axis was varied in steps of 0.1$^\circ$.
For each $\theta$, we numerically solved the Schr\"odinger equation for a $\mathrm{V_B^-}$ electron spin coupled to three neighboring $^{15}\mathrm{N}$ nuclear spins, and calculated the probabilities $p_\mathrm{sim}^{(0)}(\tau)$ and $p_\mathrm{sim}^{(1)}(\tau)$ of observing $m_S=0$ and $m_S=-1$, respectively, at readout~[See Supplementary].
Using these quantities, we expressed the PL intensity as
\begin{equation}
I(\tau) = 
\exp\left\{-\left(\frac{2\tau}{T_2}\right)^\alpha\right\}
\left(I_0 p_\mathrm{sim}^{(0)}(\tau) + I_1 p_\mathrm{sim}^{(1)}(\tau)\right)
\label{eq:fitformula_VB_Hahn}
\end{equation}
and fitted the full experimental data in \cref{fig:VBpulse_HahnEcho}(a) for each $\theta$.
The fitting parameters were $\{I_0, I_1, T_2, \alpha\}$.
We then compared the mean squared error (MSE) between the experimental data and the fitted curve for each $\theta$, and found that it was minimized at $\theta=3.8^\circ$.
We adopted the fit result at this angle and obtained $T_2=108.7\pm5.5$~ns and $\alpha=1.25\pm0.11$, as shown by the solid line in \cref{fig:VBpulse_HahnEcho}(a).

A full-waveform fit that explicitly includes the hyperfine interaction with neighboring $^{15}\mathrm{N}$ nuclear spins is important for the quantitative evaluation of $T_2$ and $\alpha$.
This is because the timescale of the ESEEM oscillation arising from the precession of the neighboring $^{15}\mathrm{N}$ nuclear spins when the electron spin state is $m_S=0$~\cite{childress2006coherent,zopes2018three} can be comparable to $T_2$~[See Supplementary].
Therefore, a fit only to the envelope may incorrectly attribute part of the signal reduction caused by ESEEM to decoherence.
In fact, as shown by the dashed line in \cref{fig:VBpulse_HahnEcho}(a), when only the envelope of the signal is fitted with
\begin{equation}
I_\mathrm{env}(\tau)
= I_\mathrm{offset}
+ C \exp\left\{-\left(\frac{2\tau}{T_2}\right)^\alpha\right\}
\label{eq:fitformula_VB_Hahn_env}
\end{equation}
we obtain $T_2=91.8\pm2.1$~ns and $\alpha=1.38\pm0.06$.
These values differ by about 16\% and 10\%, respectively, from the values $T_2=108.7\pm5.5$~ns and $\alpha=1.25\pm0.11$ obtained from the full-waveform fit based on \cref{eq:fitformula_VB_Hahn}.
Therefore, to evaluate $T_2$ and $\alpha$ more reliably, a full-waveform fit that explicitly includes the hyperfine interaction with the neighboring $^{15}\mathrm{N}$ nuclear spins is necessary.

Interestingly, the envelope-fit value $\alpha=1.38\pm0.06$ agrees well with the reported value $\alpha=1.34$ from numerical calculations for a single $\mathrm{V_{B}^-}$ coupled to nearby nuclear spins~\cite{haykal2022decoherence}. 
This agreement suggests that $T_2$ relaxation is mainly governed by a small number of local nuclear spins, and that similar relaxation dynamics appear in both single defects and ensembles. 
This is reasonable because the Fermi-contact interaction with nearby nuclear spins is strong~\cite{haykal2022decoherence,Trknyi2025}, and the relevant nuclear-spin configurations are expected to be similar for $\mathrm{V_{B}^-}$ defects at equivalent crystal sites.

To conclude, by placing an $\mathrm{h}^{10}\mathrm{B}^{15}\mathrm{N}$ thin film on a gold wire with width 1.5~$\mathrm{\mu}$m, we applied strong microwave pulses with Rabi frequencies of $f_\mathrm{R} = 0.35$--$0.6$~GHz to $\mathrm{V_{B}^-}$ and realized spin-coherence measurements with suppressed pulse errors.
As a result, from Ramsey interference, we obtained $T_2^* = 13.8 \pm 0.5$~ns with a stretch factor of $\alpha = 2$.
From Hahn echo measurements, we obtained $T_2=108.7\pm5.5$~ns and a stretch factor of $\alpha=1.25\pm0.11$.

Strongly driven pulses are also useful beyond reducing pulse errors. Faster quantum gate operations allow more pulses to be applied within the coherence time, which may improve the sensitivity of ac magnetic-field sensing using dynamical decoupling~\cite{Maze2008Nature,Taylor2008,DeLange2011,Rizzato2023}. The GHz-scale coherent dynamics observed here also suggests that $\mathrm{V_{B}^-}$ in hBN is a promising platform for exploring quantum control in the Floquet
regime beyond the rotating-wave approximation.

This study not only provides a quantitative characterization of the basic spin-coherence properties of $\mathrm{V_{B}^-}$, but also presents a standard experimental method for accurately evaluating the coherence of spin defects in two-dimensional materials with broad resonance linewidths.

\begin{acknowledgments}
We thank Mr.~Tomohiko Iijima (AIST) for the usage of AIST SCR HIM for helium ion irradiations, Dr. Toshihiko Kanayama (AIST) for helpful discussions since the introduction of HIM at AIST in 2009, and Prof. Kohei M. Itoh (Keio University) for letting us use the confocal microscope system. 
This work was partially supported by
JST, CREST Grant No.~JPMJCR23I2, Japan;
Grants-in-Aid for Scientific Research (Grants No.~JP26H02007, No.~JP25H01248, No.~JP24K21194, and New Challenge Research hosted by JSR Corporation via JSR-UTokyo Collaboration Hub, CURIE);
“Advanced Research Infrastructure for Materials and Nanotechnology in Japan (ARIM)” (Proposal No.~JPMXP1224UT1056 and No.~JPMXP1224NM0055) of the Ministry of Education, Culture, Sports, Science and Technology of Japan (MEXT); 
“World Premier International Research Center Initiative on Materials Nanoarchitectonics (WPI-MANA)” supported by MEXT;
the Mitsubishi Foundation (Grant No.~202310021);
and the Cooperative Research Project of RIEC, Tohoku University.
Y.N. acknowledges the financial support from FoPM, the WINGS Program, The University of Tokyo, and the JSPS Young Researcher Fellowship (No.~JP24KJ0692). 
\end{acknowledgments}

\bibliography{myref}

\end{document}